\documentclass[12pt,letterpaper]{article}
\usepackage[latin1]{inputenc}
\usepackage{amsmath, amsthm, amsfonts, amssymb, makeidx, graphicx, xcolor, mathrsfs, tabularx, multirow, booktabs, subcaption, lscape, url, bm}
\usepackage[spanish, english]{babel}

\usepackage[left=2.00cm, right=2.00cm, top=2.00cm, bottom=2.00cm]{geometry}

\usepackage{setspace}
\AtBeginDocument{%
	\addtolength\abovedisplayskip{0.5\baselineskip}%
	\addtolength\belowdisplayskip{0.5\baselineskip}%
	}

\usepackage[page,toc,titletoc,title]{appendix}

\usepackage[flushleft]{threeparttable}

\usepackage{fancyhdr}

\usepackage{tikz}
\usetikzlibrary{babel}
\usetikzlibrary{arrows, positioning, calc}
\tikzset{every picture/.append style={font=\footnotesize}}

\usepackage[colorlinks=true, linkcolor=black, citecolor=black, urlcolor=blue]{hyperref}

\usepackage{siunitx}
\makeindex

\usepackage{caption}
\DeclareCaptionStyle{mystyle}[margin=5mm,justification=centering]%
{font=footnotesize,labelfont=sc,margin={10mm,0mm}}
\usepackage{pgfplots}
\pgfplotsset{compat=1.16}
\usepackage{apacite}

\usepackage{doi}

\usepackage{titlesec}
\titleformat{\section}{\normalfont\Large \bfseries}{\thesection. }{0.5em}{\bfseries}
\titleformat{\subsection}{\itshape\large\bfseries}{\thesubsection.}{1em}{}
\titleformat{\subsubsection}{\itshape\normalsize\bfseries}{\thesubsubsection}{1em}{}

\begin{document}

	\renewcommand{\BOthers}[1]{et al.\hbox{}}
	\renewcommand{\BBAA}{\&}
	\renewcommand{\BBAB}{\&}

\begin{titlepage}

\selectlanguage{english}    
    \vskip 1cm
    \begin{center}
    	{\Large \bfseries Shock Symmetry and Business Cycle Synchronization:} 
    	\vskip 2pt
        {\large \bfseries Is Monetary Unification Feasible among CAPADR Countries?} 
        \vskip 1cm
        {\textbf{Author:} \emph{Jafet Baca}}
        \vskip 5pt
         \textbf{Date:} \emph{October 2021}
    \end{center}
    \vskip 0.2cm
    \begin{abstract}
    \noindent In light of the ongoing integration efforts, the question of whether CAPADR economies may benefit from a single currency arises naturally. This paper examines the feasibility of an Optimum Currency Area (OCA) within seven CAPADR countries. We estimate SVAR models to retrieve demand and supply shocks between 2009:01 - 2020:01 and determine their extent of symmetry. We then go on to compute two regional indicators of dispersion and the cost of inclusion into a hypothetical OCA for each country. Our results indicate that asymmetric shocks tend to prevail. In addition, the dispersion indexes show that business cycles have become more synchronous over time. However, CAPADR countries are still sources of cyclical divergence, so that they would incur significant costs in terms of cycle correlation whenever they pursue currency unification. We conclude that the region does not meet the required symmetry and synchronicity for an OCA to be appropiate.	
    \end{abstract}
    {\selectlanguage{spanish}
    \begin{abstract}
    \noindent Dados los esfuerzos de integración actuales, cabe preguntarse si las economías CAPARD pueden beneficiarse de un sistema de moneda única. Este documento analiza la factibilidad de una Zona Monetaria Óptima (ZMO) entre siete economías CAPARD. Se estiman modelos SVAR para extraer choques de oferta y demanda durante 2009:01 - 2020:01 y determinar su nivel de simetría. Posteriormente, se calculan dos indicadores regionales de dispersión y el costo de incorporación a una ZMO hipotética para cada país. Los resultados sugieren que los choques de naturaleza asimétrica predominan. En adición, los índices de dispersión reflejan que los ciclos económicos muestran una tendencia creciente de sincronización. Sin embargo, los países CAPARD aún constituyen fuentes de divergencia cíclica, de forma que incurrirían en costos	significativos en términos de correlación cíclica si persiguen la unión monetaria. Se concluye que la región no muestra la simetría y sincronización requeridas para que una ZMO sea apropiada.
    \end{abstract}}
    \vskip 0.2cm
    \selectlanguage{english}
    
    
    \mbox{}
\end{titlepage}
	
\thispagestyle{empty}
\tableofcontents
\newpage

\section*{Acronyms}
\thispagestyle{empty}
\begin{table}[h!]
	\begin{tabular}{ll}
		\textbf{CABEI}& Central American Bank for Economic Integration \\
		\textbf{CACM} & Central American Common Market \\
		\textbf{CAMC}& Central American Monetary Council\\
		\textbf{CAPADR}& Central America, Panama, and Dominican Republic\\
		\textbf{CAFTA-DR}& Dominican Republic-Central America-United States Free Trade Agreement\\ 
		\textbf{CPI} & Consumer Price Index \\
		\textbf{EMU}& Economic and Monetary Union of the European Union  \\
		\textbf{EU-CAAA}& European Union-Central America Association Agreement\\
		\textbf{FDI}& Foreign Direct Investment  \\
		\textbf{GDP}& Gross Domestic Product \\
		\textbf{MEAI}& Monthly Economic Activity Index \\
		\textbf{OCA}& Optimum Currency Area(s)\\
		\textbf{OLS}& Ordinary Least Squares\\
		\textbf{SAARC}& South Asian Association for Regional Cooperation\\
		\textbf{SECMCA}& Executive Secretariat of the Central American Monetary Council \\
		\textbf{SICA}& Central American Integration System\\
		\textbf{SIECA}& Central American Secretariat for Economic Integration\\
		\textbf{SIMAFIR}& System of Macroeconomic and Financial Information of the Region\\
		\textbf{SVAR}& Structural Vector Autoregression \\
		\textbf{VAR}& Vector Autoregression
	\end{tabular}
\end{table}
\newpage

\section{Introduction}

Beyond their notorious geographical proximity, Central American countries, Panama and Dominican Republic (CAPADR economies) share other gravity features, such as a common language and economically small sizes, as well as cultural and historical characteristics. In fact, Guatemala, Honduras, El Salvador, Nicaragua, and Costa Rica celebrate in 2021 the bicentenary of their independence from Spain. Not less importantly, CAPADR countries have also engaged in different free trade agreements, namely, the Dominican Republic-Central America-United States Free Trade Agreement (CAFTA-DR) and the European Union-Central America Association Agreement (EU-CAAA), among others.\\ 

Consistent with these ties, such economies have undertaken progressive efforts to achieve a higher level of integration. With the signing of the General Treaty of Central American Economic Integration in 1958 and the creation of the Central America Common Market (CACM) in 1960 as modern starting points, the region has gradually adhered to the Central American Integration System (SICA) since its foundation in 1991 \cite{Caldentey2021}. Notwithstanding, the integration process became more visible since its relaunching in 2010, achieving the most decisive step so far with the development of the Deep Integration Process (the Customs Union) among Guatemala and Honduras in 2015, and the later incorporation of El Salvador in 2018 \cite{Duran2019}. \\

Provided that the economic union and strengthening of financial systems is one of the primary purposes of SICA \cite{Castro2019}, the question of whether it is possible to move toward a gradual monetary unification arises naturally.  Such a topic is relevant insofar as adopting a common currency over other arrangements entails benefits and costs. First, participating countries experience lower transaction costs, reduced uncertainty, and increased policy confidence \cite{Zhao2009}. On the other hand, the main challenge stems from the inability to invoke independent monetary policy, which may disproportionately harm members in terms of macroeconomic stability when asymmetric aggregate disturbances prevail \cite{Padilla2021}. To the best knowledge of the authors, no previous work has explicitly addressed this issue for the CAPADR region by using an empirical background.\\

This paper seeks to shed light on the economic feasibility of an Optimum Currency Area (OCA) within CAPADR countries from the viewpoint of structural shock symmetry and business cycle synchronization. To assess the first criterion, we estimate Structural Vector Autoregression (SVAR) models on Monthly Economic Activity Index (MEAI) and Consumer Price Index (CPI) from January 2009 to January 2020 for seven CAPADR economies to retrieve aggregate supply and demand shocks. We base our identification method on long-run restrictions proposed by \citeA{Bayoumi1992}. To examine the second condition, we calculate two regional indicators of dispersion from the estimated country-specific shocks to ascertain the patterns of cyclical comovement in the spirit of \citeA{CrespoCuaresma2013}, as well as computing cost of inclusion series for each country in terms of their contribution to a higher or lower degree of business cycle synchronization within a hypothetical OCA.\\

Our results are as follows. The analysis of pairwise correlation coefficients indicates that CAPADR countries are predominantly exposed to asymmetric supply and demand shocks. In the cases in which correlations are positive and statistically significant, their magnitudes are low. Nevertheless, we identify one group of countries (Dominican Republic, Honduras, Panama, El Salvador) that may benefit partially from higher monetary coordination. In addition, our dispersion indexes depict downward trends, which provides evidence that structural shocks ---deemed as proxy series of business cycles--- have become more synchronized over time. Finally, most countries are sources of cyclical divergence, so that they would incur significant costs in case of being members of an OCA. These facts imply that CAPADR countries do not meet the required comovement features to form an OCA. \\

The rest of the paper is structured as follows. First,  we explore recent literature on the conditions used for determining an OCA and the leading empirical approaches. Next, we delve into the theoretical foundations of OCA criteria, the role of (a)symmetric shocks, and the relevance of synchronous business cycles. Consecutively, we develop the methodological considerations in the fourth section, which includes our identification framework. The fifth section presents the main estimates and results, together with their corresponding analyses. In section 6, we discuss our main findings and provide policy implications. Section 7 concludes.

\section{Literature overview}

Empirical practitioners have formulated a large part of the literature body in the realm of monetary unification on the grounds of \citeA{Mundell1961}, \citeA{McKinnon1963}, and \citeA{Kennen1969}. These pioneering contributions established that, for a group of countries to be considered an OCA, several factors must be met, including labor mobility across the region, openness and the size of the candidate economies, product diversification, and fiscal integration. Nevertheless, these criteria are not exhaustive and have been subject to sharp criticism due to their problem of inconclusiveness. Thereby, new research since the decade of 1990 has highlighted the role of trade specialization, the symmetry of macroeconomic shocks, and the similarity of financial systems, as well as the importance of production patterns among the candidate countries as indicators for a currency union \cite{Bayoumi2001}.   \\

As far as the empirical literature is concerned, researchers often follow three main approaches to assess existent monetary areas or potential groups of countries. The first line of research concentrates on evaluating the traditional OCA criteria via a descriptive and correlational analysis. The Maastricht model represents the second literature strain. This framework originally aimed at assessing EMU integration experience and evaluating if other European countries were economically qualified to join the EMU. The third methodology has focused on ascertaining the degree of symmetry of business cycles among candidate countries by quantifying underlying structural shocks.\\

Within the first framework, the analysis usually compares proxy variables capturing OCA features among the target economies. As an illustration, \citeA{Moslares2011} explore the evolution of a series of macroeconomic variables of the CACM to determine their progress towards a monetary union. The authors identify a high degree of product diversification, similarity of productive structures, and synchronicity of inflation rates. However, they also find that the GDP growth rates are weakly synchronized, as does the money supply. Furthermore, they confirm the persistence of interest rate differentials and reduced relative participation of capital inflows and inmigration from the economies. Hence, there is still a long way to go to fulfill OCA criteria.  \\

On the other hand, the Maastricht model of monetary dominance and convergence includes several thresholds on key macroeconomic fundamentals for the sake of the establishment of a currency area. For instance, for a country to be considered a potential candidate for a monetary union, it is required to have a fiscal budget deficit not greater than 3\% of GDP \cite{Hochreiter2003}. Thus, like the first approach, this methodology mainly depends on descriptive measures instead of an econometric background. Moreover, it lacks a theoretical foundation \cite{Regmi2015}. Therefore, empirical developments take into consideration the Maastricht framework more as a complementary perspective. \\

In this spirit, \citeA{Hochreiter2003} analyze the outcomes for inflation and output for Austria, Canada, the Netherlands, and New Zealand under different exchange rate configurations between 1970 and 2000. Counterfactual experiments based on SVAR models partially support the idea of a monetary union between Australia and New Zealand, a similar course of action for the Netherlands, and inflation-targeting regimes for Austria and Canada. These results notwithstanding, Australia, New Zealand, and Canada fail to meet the exchange rate criterion of the Maastricht model. Therefore, in a strict sense, these countries do not fit for a monetary union. \\

As part of their comparative analysis, \citeA{Regmi2015} examine if the eight SAARC economies (Afghanistan, Bangladesh, Bhutan, India, the Maldives, Nepal, Pakistan, and Sri Lanka) fulfill the Maastricht criteria for 1970-2011. Even though these countries have external debt amounts less than 60\% of GDP, results for the rest of the indicators are less coincident. For example, the Maldives and Pakistan experience comparatively higher inflation rates; India, the Maldives, Pakistan, and Sri Lanka display budget deficits higher than 3\%. Hence, these South Asian countries fail to meet the prerequisites for currency unification.\\

Recent applications address the OCA theory in terms of the degree of (a)symmetry of macroeconomic disturbances among potential participants. This scheme is relevant insofar as giving up an independent monetary policy to tackle macroeconomic imbalances is less costly if the candidate countries face similar shocks \cite{Padilla2021}. Indeed, highly asymmetric shocks imply a floating exchange rate might play the corrective role of shock absorber; otherwise, the exchange rate might lead to critical macroeconomic costs \cite{Hochreiter2003}, say, large fluctuations in production and output due to the disagreement on the interest rate-setting \cite{Staehr2015}.\\

In general, these studies focus on three geographical areas, namely, South Asia, the CFA Franc Zone, and Latin America, particularly South America. Meanwhile, the preferred approach to operationalize the OCA framework consists of identifying demand and supply shocks by following the strategy of \citeA{Bayoumi1992} and \citeA{Blanchard1989}. Highly correlated shocks would suggest that the economies fit for a currency unification. In this sense, \citeA{Ahn2006} explores the suitability of a deeper degree of monetary cooperation in East Asia. Given the high pairwise correlations of supply shocks estimated via an SVAR model and their statistical significance, the authors infer that seven East Asian countries\footnote{Hong Kong SAR, Indonesia, Korea, Malaysia, Singapore, Thailand, and Taiwan.} are suitable for integrating an OCA. Furthermore, the paper highlights the need for a strong political will to pursue monetary cooperation. \\

More recently, \citeA{Samba2019} evaluate the synchronization of business cycles within the CFA Franc zone for the period 1990-2013 by extracting demand and supply shocks, complemented with an indicator of dispersion. As for supply shocks, most of the correlation ratios between individual countries and the aggregate currency area turn out to be positive and relatively high. The authors find similar results for demand shocks, except for Guinea Bissau. On its part, the dispersion analysis suggests that business cycles within the CFA Franc zone are synchronized as far as demand shocks are concerned, possibly explained by the resemblance of the monetary policies conducted in the area. Meanwhile, supply shocks show signs of synchronization only after the year 2000.  \\

For the best knowledge of the author, no empirical work related to CAPADR countries has examined the economic suitability of an OCA, much less in terms of shock symmetry and business cycle synchronization. Here, the contribution of this paper. By taking this region as part of their analysis, \citeA{Hafner2018} deduce that the benefits of further monetary integration for these economies relatively outweigh the costs when compared to the rest of Latin American countries. Indeed, CACM economies benefit more from maintaining low inflation rates as it stimulates GDP and FDI flows. On the contrary, loss of autonomous monetary would induce a lack of economic stability when demand shocks or speculative attacks hit the region.

\section{Business cycle synchronization and the OCA theory}

The traditional OCA framework establishes a series of conditions that indicate whether a group of countries should pursue a monetary union, as discussed in the previous section. In the original view of \citeA{Mundell1961}, the decision of a group of countries to form a monetary area ultimately becomes a matter of weighing the advantages of removing currency conversion and the drawbacks of not being able to control country-specific shocks under an OCA. The benefits arise primarily from monetary efficiency gains ---that is, reduced transaction costs and uncertainty---, and international price convergence resulting from close integration with low-inflation countries or areas \cite{Krugman2018}. 
On the other hand, the loss of economic stability, as a consequence of leaving off monetary sovereignty, represents the main cost of adopting the single currency of an OCA.   \\

Recent approaches on OCA concentrate on policy-related issues, in particular, the nature of underlying economic shocks. Certainly, the question of whether aggregate shocks affecting potential candidates are symmetric or not is critical to forming an OCA. Indeed, the more correlated macroeconomic and sectoral disturbances are, the lower the probability of asymmetric disturbances and the cost of abandoning the independent monetary policy \cite{Cevik2014}. In concrete, a shock is regarded as asymmetric if its impact across the economies is disproportional. Under price stickiness, asymmetric shocks cause disequilibria. Specifically, countries facing negative demand shocks undergo a deflationary shock, recession, unemployment, and wage cuts, whereas the rest of the members plunge into the opposite effects \cite{Hafner2018}. \\  

\begin{figure}[t!]
	\centering
	\caption{The hypothesis of the endogeneity of OCA.}
	\includegraphics[width=0.7\linewidth]{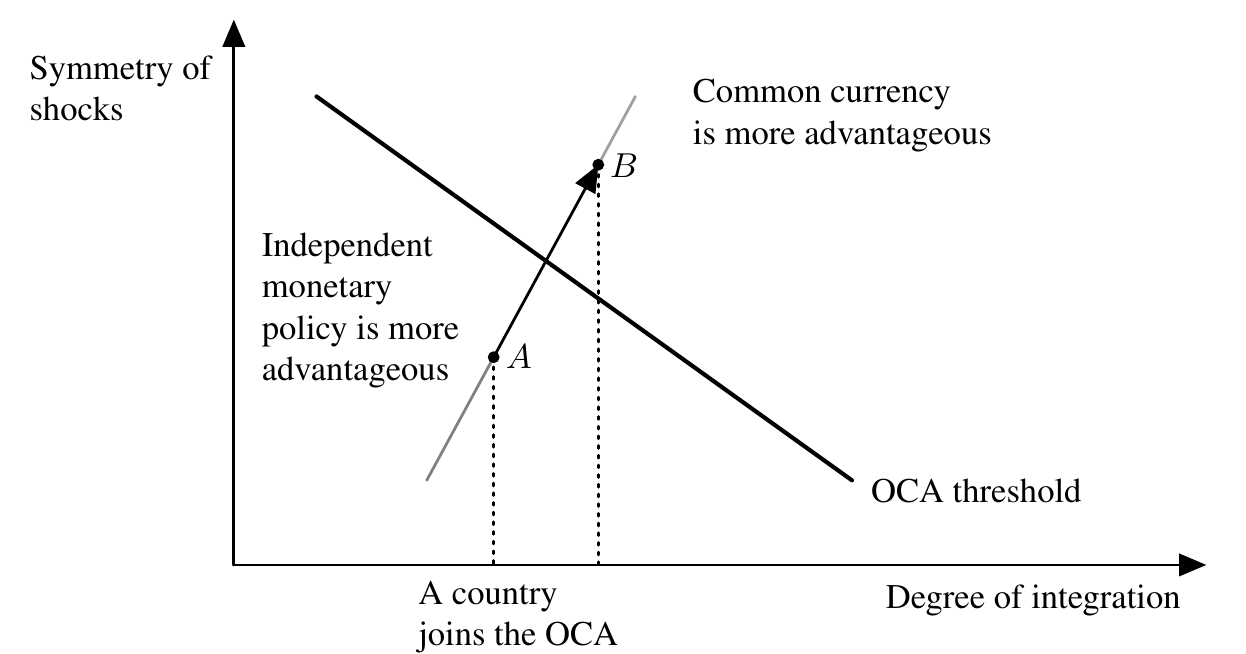}
	\label{fig:endogeneity}
	\caption*{\emph{Source.} \citeA{DeGrauwe2005}.}
\end{figure}

The formation of EMU in 1992 as the earliest experience of monetary unification revived the interest in whether an OCA requires synchronized business cycles. An alternative theoretical current asserts that business cycle correlation is likely to be endogenous (figure \hyperref[fig:endogeneity]{\ref*{fig:endogeneity}}), even in a Keynesian economy characterized by nominal rigidities and the prevalence of anti-cyclical policies \cite{Praussello2011}. The rationale goes as follows: the adoption of a single-currency system tends to increase \emph{ex-post} economic integration and intra-regional trade without prior fulfillment of OCA criteria, provided the elimination of transaction costs, riskiness, and uncertainty related to exchange rate fluctuations. Therefore, the increased intra-industry trade and facilitated FDI flows reinforce business cycle synchronization over time. In addition, the lack of idiosyncratic monetary policy also lessens the possibility of asymmetric monetary shocks and competitive devaluations, thus leading to a higher synchronization \cite{Beetsma2010}.\\

\begin{figure}[h!]
	\centering
	\caption{The specialization effect.}
	\includegraphics[width=0.7\linewidth]{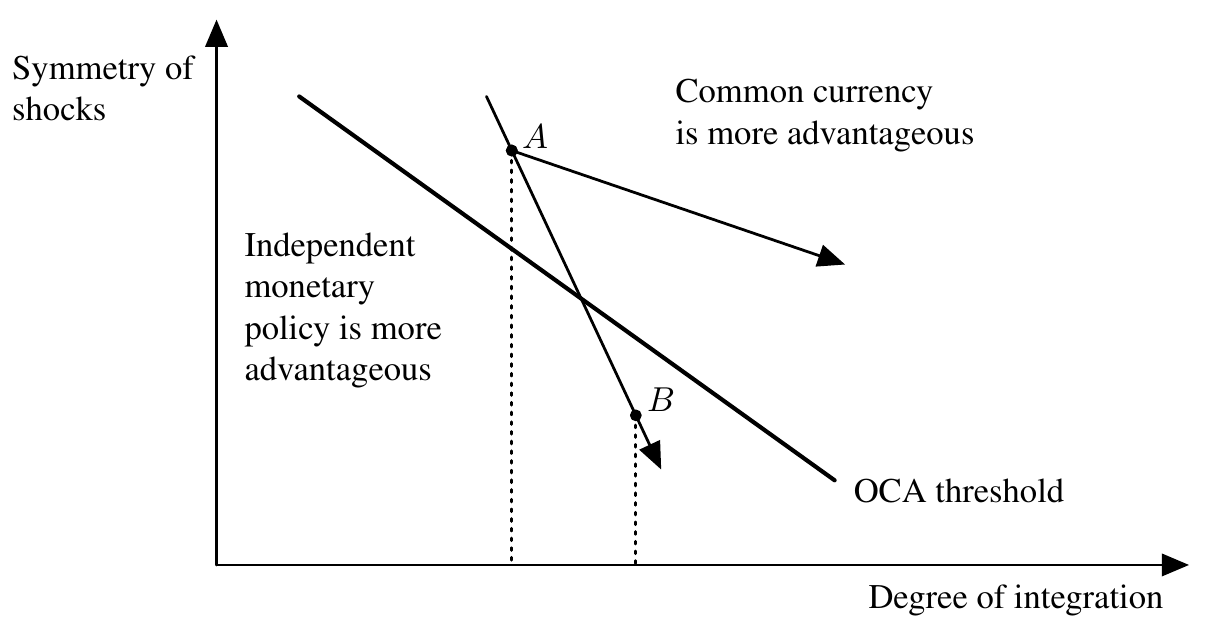}
	\label{fig:specialization}
	\caption*{\emph{Source.} \citeA{DeGrauwe2005}.}
\end{figure}

However, high trade may not necessarily result in correlated business cycles. On the contrary, candidate countries are prone to specialize in different industries in which they exhibit comparative advantage, insofar these sectors attract and agglomerate the respective regional production and exploit economies of scale (\citeNP{Mongelli2016}; \citeNP{Krugman2018}).\footnote{The so-called Krugman's specialization effect.} As a consequence, the economic structure becomes less diversified and the countries are more vulnerable to industry-specific disturbances, whose impact disseminates unequally across the OCA members. That a group of countries finds that the benefits of monetary union no longer outperform those of monetary independence depends on the relative strength of the increase in asymmetry compared to the rise in the efficiency gains of the OCA (figure \hyperref[fig:specialization]{\ref*{fig:specialization}}).  Such specialization patterns require a significant degree of labor mobility to avoid imbalances \cite{Mongelli2005}.  \\

Other explanations unveil the importance of starting economic circumstances for the implementation of an OCA. For instance, less correlated business cycles are likely to reflect significant dissimilarities in economic structure, which makes it harder for countries to achieve the desired optimality \cite{Botto2018}. Furthermore, the existence of significant development gaps in the context of an initial monetary union may be detrimental for least developed members. If the central monetary authority aims at low inflation rates, the latter may experience a distorted investment structure, excessive demand, and loss of competitiveness that will prompt them to undergo a long-standing deflation to recover \cite{Dellas2009}. \\

Therefore, assessing business cycle synchronization and the symmetry of shocks among Central American countries is crucial to reckon their feasibility for monetary integration. In the absence of political commitment and working adjustment mechanisms to shocks such as free labor mobility and fiscal transfers dictated by a unified fiscal authority, asymmetric shocks may disproportionately impact developing countries in the region. Thereby, a hypothetical premature OCA can prevent these countries from achieving a higher level of economic development.

\section{Data and empirical methodology}

\subsection{Empirical specification}

To start, we build upon the standard model of aggregate supply and aggregate demand  specified by \citeA{Bayoumi1992}. Such an approach supposes that the aggregate demand curve has a negative slope in the price-output system of coordinates. In contrast, the short-run aggregate supply curve is upward-sloping. In the long run, the aggregate supply curve is vertical. Under this scenario, a positive supply shock embodies a permanent effect on output, while a positive demand disturbance, in accordance with the natural rate hypothesis, affects output temporarily. In addition, a positive demand shock increases permanently the price level, whereas an outward shift in the aggregate supply curve disturbance reduces it.\\ 

Following the strategy outlined by \citeA{Bayoumi1992}, let us consider a system which approximates the underlying true model by an infinite moving average of a vector of variables (say, $X_t$) and an equivalent number of independent white-noise disturbances $\varepsilon_t$; in other words
\begin{equation} \label{eq1}
	X_t = A_0 \varepsilon_t + A_1 \varepsilon_{t-1} + A_2 \varepsilon_{t-2} + \cdots = \sum_{i=0}^{\infty} L^i A_i \varepsilon_t
\end{equation} 
where $L$ represents, as usual, the lag operator, and the matrices $A_i$ correspond to the impulse-response functions of the shocks to the elements of $X_t$. Letting $X_t$ be the vector comprising the changes in the logarithms of output ($\Delta y_t$) and prices ($\Delta \pi_t$), the model (\ref{eq1}) transforms into:
\begin{equation} \label{eq2}
	\left[\begin{array}{c}
			\Delta y_t \\ \Delta \pi_t
		\end{array}\right]
	= \sum_{i=0}^{\infty} L^i \left[\begin{array}{cc}
		a_{11i} & a_{12i} \\ a_{21i} & a_{22i}
	\end{array}\right] \left[\begin{array}{c}
	\varepsilon_{st}\\ \varepsilon_{dt}
\end{array}\right]
\end{equation}
in which $\varepsilon_{d}$ and $\varepsilon_{s}$ are independent demand and supply shocks with normalized variances so that $\mathrm{var}(\varepsilon_d) = \mathrm{var}(\varepsilon_{s})= 1$. On the other hand, our initial assumption on the  transitory influence of demand shocks over output translates into the following restriction:
\begin{equation} \label{eq3}
	\sum_{i=0}^{\infty} a_{12i} = 0
\end{equation}

Given that the two variables are stationary, we can recover the structural shocks entailed in equations (\ref{eq1}) and (\ref{eq2}) by estimating a reduced-form Vector Autoregression (VAR) model and converting it into its bivariate moving average representation as:
    \begin{equation}\label{eq4}
    	\begin{split}
    	X_t & = B_1 X_{t-1} + B_2 X_{t-2} + \cdots +B_n X_{t-n} + \epsilon_t \\
    	& = \left[I - B(L)\right]^{-1}\epsilon_t \\
    	& = \left(I + B(L) + B(L)^2 + \cdots \right)\epsilon_t \\
    	& = \epsilon_t + D_1 \epsilon_{t-1} + D_2 \epsilon_{t-2} + \cdots 
    \end{split}
    \end{equation}
   
where $\epsilon_t = [\begin{array}{cc}
	\epsilon_{st} & \epsilon_{dt}
\end{array}]^\top$, being $\epsilon_{dt}$ and $\epsilon_{st}$ the (possibly correlated) VAR residuals. A direct comparison of equations (\ref{eq2}) and (\ref{eq4}) reveals  the existence of a conversion factor between the structural shocks and the VAR residuals, namely: 
\begin{equation}\label{eq5}
 \left[\begin{array}{c}
 	\epsilon_{st}\\ \epsilon_{dt}
 \end{array}\right]	= \left[\begin{array}{cc}
 a_{110} & a_{120} \\ a_{210} & a_{220}
\end{array}\right] \left[\begin{array}{c}
\varepsilon_{st} \\ \varepsilon_{dt}
\end{array}\right] \Longrightarrow \epsilon_t = A_0\varepsilon_t
\end{equation}

therefore, the identification of the demand and supply shocks only requires computing the four elements of $A_0$. In turn, we need four restrictions to recover these parameters. The simple normalizations $\mathrm{var}(\varepsilon_{dt}) = \mathrm{var}(\varepsilon_{st}) = I$ provide us with two of them. The assumption of orthogonality between demand and supply shocks yields a third restriction. The fourth restriction comes from the premise that demand disturbances exhibit temporary effects on output. In accordance with the VAR stated above, this implies that:
\vskip 5pt
\begin{equation}\label{eq6}
	\sum_{i=0}^{\infty} \left[\begin{array}{cc}
		d_{11i} & d_{12i} \\
		d_{21i} & d_{22i}
	\end{array}\right] \left[\begin{array}{cc}
		C_{11} & C_{12}\\
		C_{21} & C_{22}
	\end{array}\right] = \left[\begin{array}{cc}
		* & 0 \\
		* & *
	\end{array}\right]
\end{equation}
\vskip 10pt
where $*$ is a placeholder. With the restrictions previouly stated, we can get the unique matrix $A_0$, and consequently the demand $\left(\varepsilon_{dt}\right)$ and supply shocks $\left(\varepsilon_{st}\right)$ series, as desired.

\subsection{Data and macroeconometric analysis}

The current analysis focuses on seven CAPADR countries: Costa Rica (CRI), Dominican Republic (DOM), Guatemala (GTM), Honduras (HND), Nicaragua (NIC), Panama (PAN), and El Salvador (SLV), all of them members of SICA. We perform the estimation of the reduced-form VAR models via Ordinary Least Squares (OLS) using monthly data  from January 2009 to January 2020 (a total of 133 observations).\footnote{Provided that the central purpose of the paper is not forecasting the evolution of MEAI and CPI, we follow \citeA{Lenza2020} in ruling out data points after January 2020 given the context of the COVID-19 pandemic, as well as figures before 2009 corresponding to the onset of the 2008 financial crisis. It is worth highlighting that the earliest date one finds complete data for all CAPADR countries is January 2007.} We use MEAI and CPI for each country as proxy variables of GDP and the price level. Data is sourced from the System of Macroeconomic and Financial Information of the Region (SIMAFIR) of the Central American Monetary Council (CAMC).\footnote{Quarterly real GDP data is available for four economies. Dominican Republic and El Salvador report Volume Indexes instead. As a result of the lack of comparability and consistency, the paper employs MEAI data. Yet imperfect, we consider this index sheds light on the trend of economic activity and thus enables us to examine the correlation of shocks.} We normalize all variables to 2010 through the usual proportionality technique, transform the series into their natural logarithm values, and adjust them seasonally through the X-13 ARIMA-SEATS program of the United States Census Bureau. \\  

\begin{figure}[h!]
	\centering
	\caption{Monthly Economic Activity and Consumer Price Indexes of CAPADR countries.}
	\includegraphics[width=1\linewidth]{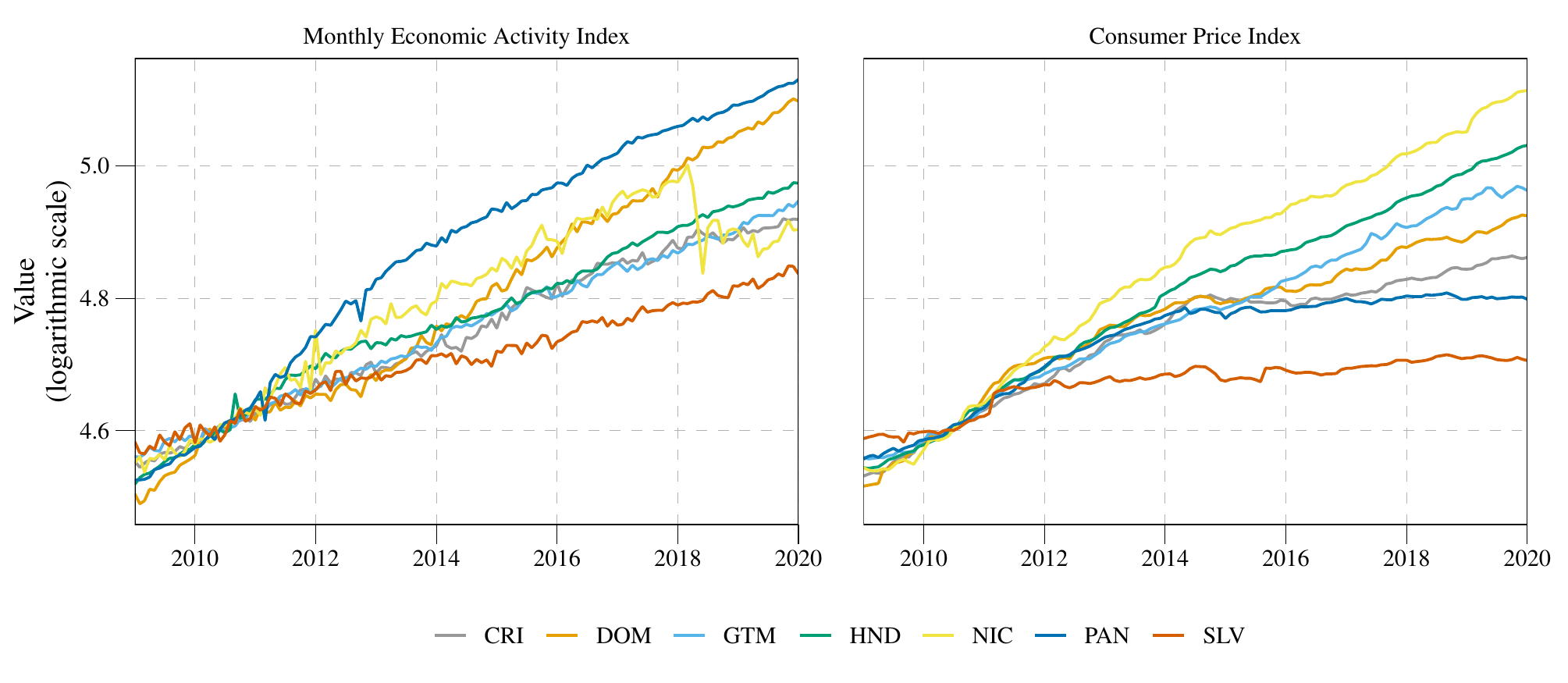}
	\caption*{\emph{Source.} Own elaboration based on CAMC data.}
\end{figure}

Before estimating the VAR models, it is crucial to examine the data-generating process of the series and determine if country-specific MEAI and CPI variables are cointegrated. According to the results of the Augmented Dickey-Fuller test, all series correspond to $I(1)$ processes at the 5\% significance level (see Table \ref{t1}). Likewise, both trace and maximum eigenvalue statistics of the Johansen cointegration test reject their null hypothesis at the same significance level (see Table \ref{ta1}). Therefore, we estimate the VAR models by including the first difference of the logarithm of the variables,\footnote{Thus, we can interpret the results in terms of MEAI growth rate and CPI inflation.} tailored to the specification of equation (\ref{eq4}):

\begin{equation}\label{eq7}
	X_t = \sum_{i=1}^{p} B_iX_{t-i} + u_t
\end{equation}
\vskip 10pt
where $X_t = [\begin{array}{cc}\Delta \ln \mbox{MEAI}_t & \Delta \ln \mbox{CPI}_t \end{array}]^\top$.\\

As for the selection of the optimal lag structure, we follow a sequential approach. Indeed, whenever the specification suggested by Akaike, Schwarz, and Hannan-Quin information criteria does not satisfy the required econometric properties, we include additional lags. In order to avoid excessive consumption of degrees of freedom, we set the maximum number of lags equal to 12. As a result, the effective lag lengths range from 7 to 9 lags. All models meet the stability condition in that the roots of their characteristic polynomials lie within the unit circle, and the residuals do not show either serial correlation or heteroskedasticity.\footnote{We also apply a CUSUM test to assess stability. Whenever the diagnostic shows a structural break, we invoke the \citeA{Bai2003} test to identify the exact break date for MEAI and CPI variables, and include dummy variables accordingly. As a result, we introduce dummy variables for the VAR models of Nicaragua (MEAI: 2018M4, CPI: 2014M8) and El Salvador (MEAI: 2012M6, CPI: 2011M8).} After the obtention of these models, we use \citeA{Blanchard1989} decomposition method to retrieve the aggregate supply and demand shocks, as well as their sizes and adjustment speeds in response to those shocks for each of the economies into consideration.

\begin{table}[t!]
	\centering
	\caption{Results of the Augmented Dickey-Fuller unit root test}
	\label{t1}
	\footnotesize
	\begin{threeparttable}
	\begin{tabular}{@{}l*{2}{S[table-format=+1.3, table-space-text-post = {$^{***}$}]}cl*{2}{S[table-format=+1.3, table-space-text-post = {$^{***}$}]}c@{}}
		\toprule \toprule
		\multirow{2}{*}{Country} & \multicolumn{3}{c}{MEAI} &  &   \multicolumn{3}{c}{CPI}  \\ \cmidrule{2-4} \cmidrule{6-8}
		& \multicolumn{1}{c}{Level} & \multicolumn{1}{c}{First diff.}  & \multicolumn{1}{c}{Conclusion}  & & \multicolumn{1}{c}{Level}  & \multicolumn{1}{c}{First diff.} &  \multicolumn{1}{c}{Conclusion} \\ \midrule 
		CRI & -1.443 & -5.678$^{***}$ & $I(1)$ & & -1.814 & -4.108$^{***}$ & $I(1)$ \\
		DOM & -1.406 & -5.050$^{***}$ & $I(1)$ & & -2.460 & -4.113$^{***}$ & $I(1)$ \\
		GTM & -2.935 & -7.164$^{***}$ & $I(1)$ & & -1.988 & -4.295$^{***}$ & $I(1)$ \\
		HND & -2.365 & -4.269$^{***}$ & $I(1)$ & & -1.328 & -4.153$^{***}$ & $I(1)$ \\
		NIC & -0.400 & -5.970$^{***}$ & $I(1)$ & & -1.461 & -5.124$^{***}$ & $I(1)$ \\
		PAN & -0.304 & -4.877$^{***}$ & $I(1)$ & & -0.626 & -3.640$^{**}$ & $I(1)$ \\
		SLV & -2.725 & -5.999$^{***}$ & $I(1)$ & & -1.670 & -4.841$^{***}$ & $I(1)$  \\ \bottomrule
	\end{tabular}
    \begin{tablenotes}
    	\footnotesize
    	\item[a]\emph{Note.} The symbols $^*$, $^{**}$, and $^{***}$ denote rejection of the null hypothesis of unit root at the 10\%, 5\%, and 1\% significance levels, respectively. Specification with a linear trend.
    	\item[ ]\emph{Source.} Own calculation.
    \end{tablenotes}
\end{threeparttable}
\end{table}

\subsection{Analysis of dispersion and the cost of inclusion in an OCA}

In order to exploit the time variability of the data and as a complement to the correlation analysis between supply and demand shocks, this paper also aims at measuring the degree of business cycle synchronicity among CAPADR countries. In this spirit, we employ the indicator of dispersion constructed by \citeA{CrespoCuaresma2013}, which consists of the cross-country standard deviation series of structural shocks weighted by their economic size:
\begin{equation}\label{eq8}
	\widehat{S}_t = \left[\sum_{i=1}^{N}\omega_{it}\left(\widehat{x}_{it} - \sum_{j=1}^{N}\omega_{jt}\widehat{x}_{jt}\right)^2\Bigg/\left(1-\sum_{i=1}^{N}\omega^2_{it}\right)\right]^{\frac{1}{2}}
\end{equation}\\
where $\widehat{x}_{it}$ denotes the demand or supply shock series, $N$ the number of countries, and $\omega_{it}$ the respective weight. As per \citeA{CrespoCuaresma2013} and \citeA{Samba2019}, we consider that the structural shock behavior approximates business cycle movements. Following \citeA{Cardoza2021}, we employ GDP, Power Purchasing Parity data from the World Bank for the construction of these weights (see Table \ref{ta2}). Since we can define convergence as a decrease of the standard deviation of the structural shocks across the countries considered, a smaller value of $\widehat{S}_t$ implies a higher symmetry of shocks and a greater level of synchronization of business cycles among candidate countries \cite{Samba2019}. \\

On the other hand, the incorporation of a country in a (new or already established) currency area may lead to a higher degree of synchronicity of business cycles when the endogeneity hypothesis holds. However, supply and demand shocks can become less symmetric over time if, on the contrary, a specialization process occurs. In line with \citeA{CrespoCuaresma2013}, we calculate the cost of inclusion for each country in the monetary area as the influence of each economy on the degree of business cycle synchronization at a given span as follows:

\begin{equation}\label{eq9}
	C_{t,i|\mathcal{G}}= \dfrac{\widehat{S}_{t|\mathcal{G}-i} - \widehat{S}_{t|\mathcal{G}}}{\widehat{S}_{t|\mathcal{G}}}
\end{equation}

in which $\widehat{S}_{t|\mathcal{G}}$ stands for the indicator of dispersion computed for all the countries of the group $\mathcal{G}$, whereas $\widehat{S}_{t|\mathcal{G}}$ indicates the cross-country standard deviation series for the group $\mathcal{G}$ excluding the country $i$. In other words, we quantify the cost of inclusion as the rate of change in our dispersion index. The cost series $C_{t,i|\mathcal{G}}$ takes positive values whenever the introduction of the country $i$ in the OCA reduces the (weighted ) standard deviation of the structural shocks, which is also a sign of convergence. Conversely, it takes negative values if the new member induces an increase of the standard deviation (hence leading to divergence) of $\mathcal{G}$. In the current analysis, we recalculate the dispersion index for the seven subsets of size $6$ in order to get the cost series for each CAPADR economy. 

\section{Findings and discussion}

\subsection{Correlation of macroeconomic shocks}

Table \ref{t2} summarizes the Pearson correlation coefficients of estimated aggregate supply shocks for each pair of CAPADR countries from October 2009 to January 2020. Most of the coefficients reported are positive but lack statistical significance, which reflects that supply-side shocks within CAPADR countries are generally asymmetric.\footnote{Following \citeA{Ahn2006}, we categorize shocks as symmetric if their sign is positive and significantly different from zero. Otherwise, we deem them as asymmetric.} The small sizes of pairwise correlations provide additional evidence of the weak comovement of shocks. Remarkably, shock correlation turns out to be highest between Nicaragua and Costa Rica, and El Salvador and Guatemala, which is consistent given that distance works as a robust gravity variable \cite{Baxter2005}. The correlation patterns do not support either the idea of a general OCA or a smaller one comprising at least 3 countries.\footnote{We consider that a subset of $n\geq 3$ economies may find monetary unification advantageous if each country presents positive and significant correlations at the 5\% significance level with at least other $n-1$ countries.}  \\

\begin{table}[h!]
	\centering
	\caption{Correlation matrix of supply shocks, 2009-2020.}
	\label{t2}
	\footnotesize
	\begin{threeparttable}
		\begin{tabular}{@{}l*{7}{S[table-format=+1.3, table-space-text-post = {$^{***}$}]}@{}}
			\toprule \toprule
			Country & \multicolumn{1}{c}{CRI} & \multicolumn{1}{c}{DOM} & \multicolumn{1}{c}{GTM} & \multicolumn{1}{c}{HND} & \multicolumn{1}{c}{NIC} & \multicolumn{1}{c}{PAN} & \multicolumn{1}{c}{SLV} \\ \midrule
			CRI & 1.000 &  &  &  &  &  &  \\
			DOM & 0.140 & 1.000 &  &  &  &  &  \\
			GTM & 0.051 & 0.177$^{*}$ & 1.000   &  &  &  &  \\
			HND & 0.031 & -0.194$^{**}$ & 0.101 & 1.000 &  &  &  \\
			NIC & 0.335$^{***}$ & 0.103 & 0.198$^{**}$ & 0.228$^{**}$ & 1.000  &  &  \\
			PAN & 0.074 & 0.080 & -0.085 & -0.037 & 0.067 & 1.000  &  \\
			SLV & -0.063 & 0.139 & 0.269$^{***}$ & 0.085 & 0.074 & -0.241$^{***}$ & 1.000  \\ \bottomrule
		\end{tabular}
		\begin{tablenotes}
			\footnotesize
			\item[a]\emph{Note.} Symbols $^*$, $^{**}$, and $^{***}$ denote statistical significance at the 10\%, 5\%, and 1\% levels, respectively.
			\item[ ]\emph{Source.} Own calculation.
		\end{tablenotes}
	\end{threeparttable}
\end{table}

Interestingly, two pairs of countries ---Panama and El Salvador, and Dominican Republic and Honduras--- depict significant negative relationships between their supply shocks series. Empiric literature points out differences in economic structure and labor market institutions, lack of fiscal discipline, as well as dissimilarities in trade specialization patterns as some possible drivers for such an outcome \cite{Samba2019}.  It is worth mentioning that Panama presents the lowest correlations with the rest of the countries (-0.027 on average), possibly due to its more diversified economic structure.\\

Concerning demand shocks, the majority of pairwise correlations are positive; however, few are statistically different from zero. Thus, it is possible to conclude that demand shocks fail to show the required symmetry to consider that an immediate or short-run process of monetary unification is feasible. In addition, the modest correlations confirm the heterogeneity of CAPADR economies as far as monetary and fiscal policies are concerned. We identify one country arrangement that may benefit from closer integration per the results of Table \ref{t3}: Dominican Republic, Panama, Honduras, and El Salvador (G1). Nevertheless, their low correlation magnitudes confirm that there is still a long way to go before considering suitable the adoption of a formal OCA within these countries.\\

\begin{table}[h!]
	\centering
	\caption{Correlation matrix of demand shocks, 2009-2020.}
	\label{t3}
	\footnotesize
	\begin{threeparttable}
	\begin{tabular}{@{}l*{7}{S[table-format=+1.3, table-space-text-post = {$^{***}$}]}@{}}
		\toprule \toprule
		Country & \multicolumn{1}{c}{CRI} & \multicolumn{1}{c}{DOM} & \multicolumn{1}{c}{GTM} & \multicolumn{1}{c}{HND} & \multicolumn{1}{c}{NIC} & \multicolumn{1}{c}{PAN} & \multicolumn{1}{c}{SLV} \\ \midrule 
		CRI & 1.000 &  &  &  &  &  &  \\
		DOM & 0.197$^{**}$ & 1.000 &  &  &  &  &  \\
		GTM & 0.193$^{**}$ & 0.119 & 1.000 &  &  &  &  \\
		HND & 0.130 & 0.279$^{***}$ & 0.163$^{*}$ & 1.000 &  &  &  \\
		NIC & 0.026 & -0.048 & 0.178$^{**}$ & -0.091 & 1.000 &  &  \\
		PAN & 0.066 & 0.242$^{***}$ & 0.003 & 0.096 & -0.066   & 1.000  &  \\
		SLV & 0.047 & 0.223$^{**}$ & 0.251$^{***}$ &  0.227$^{**}$ & 0.051 & 0.272$^{***}$ & 1.000  \\ \bottomrule
	\end{tabular}
\begin{tablenotes}
	\footnotesize
	\item[a]\emph{Note.} Symbols $^*$, $^{**}$, and $^{***}$ denote statistical significance at the 10\%, 5\%, and 1\% levels, respectively.
	\item[ ]\emph{Source.} Own calculation.
\end{tablenotes}
\end{threeparttable}
\end{table}

Before we move on to the analysis of the dispersion of disturbances and cost of inclusion, we reckon the size of aggregate shocks and the speed of adjustment as complementary indicators of the feasibility of CAPADR countries to form an OCA. Indeed, the smaller the shocks that a group of countries experiences and the faster the adjustment in response to disturbances, the less significant the impact of asymmetric shocks and the more favorable such economies qualify for a currency union \cite{Ahn2006}. To do so, we perform an impulse-response analysis. We gauge the size of demand and supply shocks as the long-run MEAI and CPI effects, respectively. Meanwhile, we measure the speed of adjustment by the first twelve-month response as a share of the long-run impacts.\\

\begin{table}[h!]
	\centering
	\caption{Size of shocks and speed of adjustment, 2009-2020.}
	\label{t4}
	\footnotesize
	\begin{threeparttable}
		\begin{tabular}{@{}lccccc@{}}
			\toprule \toprule
			\multicolumn{1}{l}{\multirow{2}{*}{Country}} & \multicolumn{2}{c}{Supply shocks} &  & \multicolumn{2}{c}{Demand shocks} \\ \cmidrule{2-3} \cmidrule{5-6} 
			\multicolumn{1}{c}{} & \multicolumn{1}{c}{Size} & \multicolumn{1}{c}{Adjustment speed} &  & \multicolumn{1}{c}{Size} & \multicolumn{1}{c}{Adjustment speed} \\ \midrule 
			CRI & 0.021 & 0.383 &  & 0.007 & 0.585 \\
			DOM & 0.051 & 0.235 &  & 0.010 & 0.854 \\
			GTM & 0.025 & 0.339 &  & 0.004 & 0.612 \\
			HND & 0.045 & 0.159 &  & 0.006 & 0.600 \\
			NIC & 0.078 & 0.132 &  & 0.005 & 0.212 \\
			PAN & 0.060 & 0.155 &  & 0.005 & 0.620 \\
			SLV & 0.007 & 0.805 &  & 0.004 & 0.821 \\ \cmidrule{1-6}
			Average & 0.041 & 0.316 & & 0.006 & 0.615 \\ \bottomrule
		\end{tabular}
		\begin{tablenotes}
			\footnotesize
			\item[]\hspace{-2pt}\emph{Source.} Own calculation.
		\end{tablenotes}
	\end{threeparttable}
\end{table}

Table \ref{t4} reports the calculations. On average, the CAPADR countries experience relatively similar-sized demand shocks. In contrast, supply shocks show greater variability.  It is important to note that, on average, the group G1  we identified in line with their cross-country correlations shows a greater size of supply shocks (0.054) than the whole sample. In addition, this grouping face more similar-sized demand shocks (0.006) than the regional average. As for the speed of adjustment, approximately one-third of the regional change in MEAI in response to a supply shock occurs within one year, while almost two-thirds of inflation adjustment takes place within the same period. On average, Dominican Republic, Honduras, Panama, and El Salvador adjust faster (0.451) to a unit-change in supply disturbances than the countries altogether.  On the other hand, these economies adjust at a more rapid (0.724) and than the CAPADR region to a demand-side shock. However, the rapid response of El Salvador for both demand and supply shocks might drive these results.

\subsection{Comovements of shocks and the cost of inclusion}

\begin{figure}[h!]
	\centering
	\caption{Indicators of dispersion for the CAPADR region.}
	\label{f4}
	\includegraphics[width=1\linewidth]{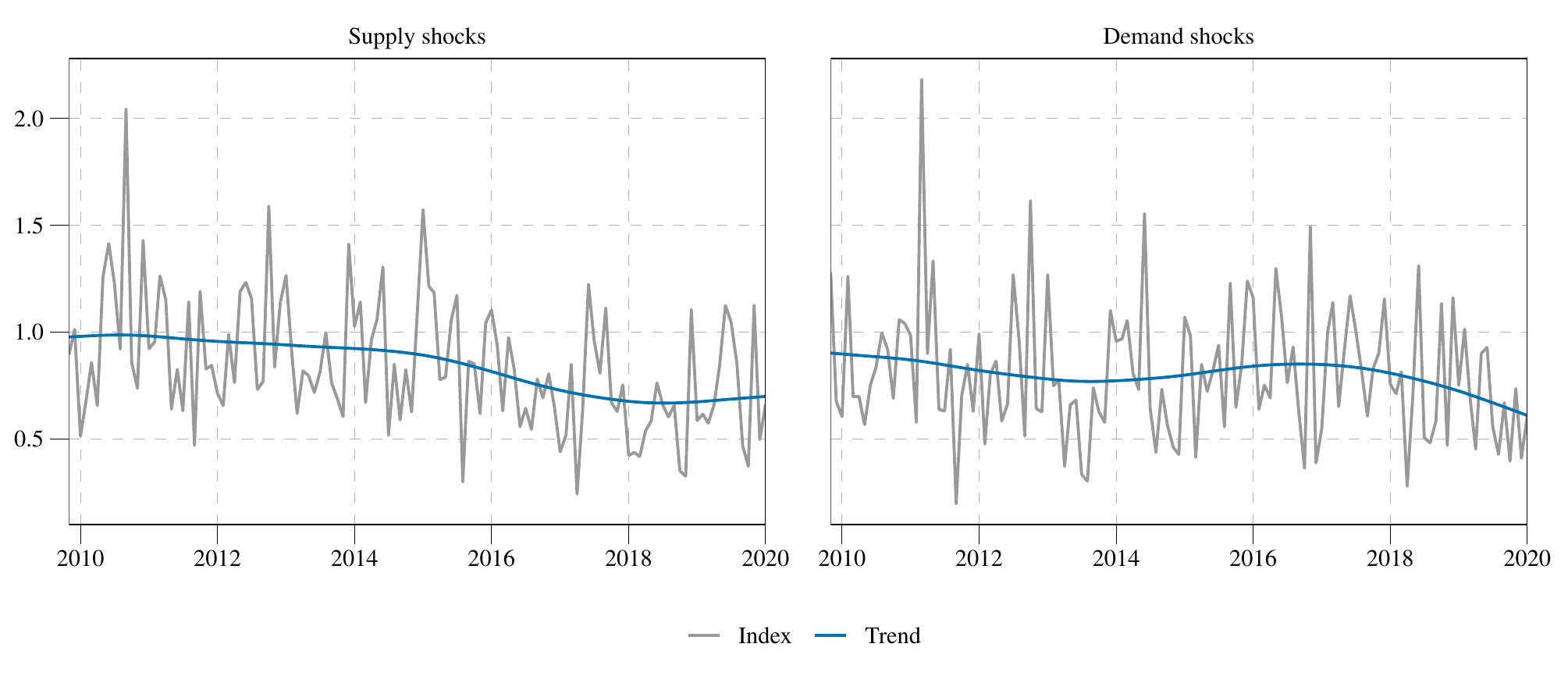}
	\caption*{\emph{Source.} Own elaboration.}
\end{figure}

In this section, we delve into the dynamics of the weighted cross-country standard deviation series, as captured by our indicator of dispersion formulated in equation (\ref{eq8}). Figure \ref{f4} displays the evolution of the respective indexes for supply and demand disturbances between October 2010 and January 2020, together with their trends computed by using the \citeA{Hodrick1997} filter. By first examining the series related to supply shocks, we can infer that the region has experienced a convergence process during the study period, except for the first and last years into consideration. Indeed, the dispersion index of supply disturbances shows a trend decrease of 28.5\% between October 2019 and January 2020. In other words, CAPADR countries show a tendency toward synchronized supply shocks.\\

With regard to the dispersion of aggregate demand shocks, the qualitative conclusions are alike. Figure \ref{f4} depicts a general downward trend, thus providing evidence that demand disturbances become more synchronized over time. Indeed, the trend value corresponding to January 2020 is 32.4\% less than its October 2010 figure. However, it is possible to identify two convergence periods (2009-2013 and 2017-2020) and one intermediate divergence period (2014-2017). The overall evolution of the dispersion of demand shocks is remarkable since the coordination of macroeconomic policies among CAPADR countries is still weak \cite{Alvarado2019}. \\

Table \ref{t5} presents the values of the cost of inclusion for each country, also calculated separately for supply and demand shocks as specified in equation (\ref{eq4}). To facilitate the discussion, we split our analysis span and report the measures at three periods: January 2010, January 2015, and January 2020. The picture is neat. The seven CAPADR economies constitute sources of cyclical divergence; in other words, the incorporation of one of these countries into a currency union integrated by the remaining six nations would increase the dispersion of shocks and consequently reduce the degree of business cycle synchronization of the whole OCA. Nevertheless, such an effect is slightly smaller in January 2020 than the other two periods. On their parts, Nicaragua and El Salvador appear as the main drivers of regional divergence in supply shocks in January 2010. Conversely, Dominican Republic and Guatemala perform this role in January 2020.  \\

Considering demand disturbances, Guatemala, Nicaragua, and Honduras show the highest cost of inclusion in comparison to the rest of the economies in January 2010. For the second date, all countries display a decrease or a negligible increase with respect to the first cut-off, except for Panama, which turns out to be the predominant trigger of regional demand shock divergence. During the last sample date, Costa Rica and Guatemala own the greatest misalignments when compared to the rest of the countries. Interestingly, the development of demand shocks in Dominican Republic, yet exiguous, induce the unique sign of cyclical convergence within the months considered. On average, cost series for both supply and demand shocks present a slight rise in January 2015 and a greater decay in January 2020.\\

\begin{table}[h!]
	\centering
	\caption{Cost of inclusion into an OCA.}
	\label{t5}
	\footnotesize
	\begin{threeparttable}
		\begin{tabular}{@{}l*{3}{S[table-format=+1.4]}c*{3}{S[table-format=+1.4]}@{}}
			\toprule \toprule
			\multicolumn{1}{l}{\multirow{2}{*}{Country}} & \multicolumn{3}{c}{Supply shocks} &  & \multicolumn{3}{c}{Demand shocks} \\ \cmidrule{2-4} \cmidrule{6-8} 
			\multicolumn{1}{c}{} & \multicolumn{1}{c}{2010M1} & \multicolumn{1}{c}{2015M1} & \multicolumn{1}{c}{2020M1} &  & \multicolumn{1}{c}{2010M1} & \multicolumn{1}{c}{2015M1} & \multicolumn{1}{c}{2020M1} \\ \midrule
			CRI & -0.043 & -0.276 & -0.092 &  & -0.075 & -0.065 & -0.324 \\
			DOM & -0.059 & -0.110 & -0.259 &  & -0.054 & -0.072 & 0.005 \\
			GTM & -0.031 & -0.068 & -0.163 &  & -0.164 & -0.022 & -0.159
			\\
			HND & -0.026 & -0.017 & -0.048 &  & -0.155 & -0.025 & -0.020 \\
			NIC & -0.283 & -0.002 & -0.004 &  & -0.116 & -0.036 & -0.079 \\
			PAN & -0.031 & -0.171 & -0.072 &  & -0.091	& -0.685 & -0.005   \\
			SLV & -0.251 & -0.097 & -0.029 &  & -0.041 & -0.005 & -0.032  \\ \cmidrule{1-8}
			Average & -0.103 & -0.106 & -0.095 &  & -0.100 & -0.130 & -0.088   \\ \bottomrule
		\end{tabular}
		\begin{tablenotes}
			\footnotesize
			\item[]\hspace{-2pt}\emph{Source.} Own calculation.
		\end{tablenotes}
	\end{threeparttable}
\end{table}

It is also of interest to discuss the results regarding the cost of inclusion in light of the country group identified in the previous section. Dominican Republic, Honduras, Panama, and El Salvador average -0.092 as to supply shocks in the first period, and -0.099 and -0.102 for the other two dates. Concerning demand shocks, G1 economies go from -0.085 in January 2010 to -0.197 and -0.013 in January 2015 and 2020, respectively, with the latter comparing favorably with the regional average. Hence, the subregion performs better than CAPADR in terms of costs toward the end of the study period when we consider demand shocks. This implies that each member induce a lower increase in the dispersion levels of the group than a new member in the whole region, on average. The result regarding supply shocks is the opposite.  \\

From the above analysis, we conclude that asymmetric supply and demand shocks prevail in the CAPADR countries in light of the low pairwise correlation coefficients. Even though the indicators of dispersion for both types of disturbances present a tendency toward more synchronized business cycles, such behavior does not translate accordingly into a significant decrease in the cost of inclusion. Moreover, CAPADR countries are drivers of cyclical divergence, so that they all incur substantial costs and lead to higher dispersion when incorporated into a hypothetical monetary union. In summary, an OCA is still unfeasible for CAPADR countries.

\section{Discussion and policy implications}

The findings of the previous section provide evidence that  CAPADR economies have a long way to go to ensure cyclic convergence and become an OCA. While the integration process is still heading toward a Customs Union and a formal process of a monetary union does not take part in the current integration treaties \cite{Alvarado2019}, we consider that the empirical evidence of this paper may work as a benchmark which future research on this topic may refer to in case a further-reaching integration process takes place.\\

Our results are partly similar to those presented by previous approaches to the possibility of higher regional integration. Indeed,  \citeA{Moslares2011} conclude that CACM countries partially meet the necessary optimality conditions for a monetary union to be successful. In particular, the authors compute a reduced level of synchronization of GDP growth across the five nations. On its part, \citeA{Alvarado2019} asserts that current macroeconomic conditions are not appropriate for currency unification, given that multilateral monetary policy might not fit the specific needs of some economies in case of country-specific disequilibria.\\

On the other hand, providing some explanations behind the predominance of asymmetric shocks, the downward trends of the dispersion indexes, and the negative costs of inclusion is, to a certain extent, speculative. It is worth clarifying that the identification of specific determinants of business cycle synchronization is beyond the scope of the paper. Nevertheless, the empirical literature has identified several underlying factors that may trigger cycle correlation. For instance, \citeA{Baxter2005} infer that bilateral trade and distance are positively and negatively related to business-cycle comovements, respectively. Keeping this idea in mind, \citeA{Giovanni2010} demonstrate that vertical production linkages account for almost one-third of the effect of bilateral trade on business cycle correlation. Similar production structure also play a positive role in synchronization \cite{Dees2012}.\\

Under these bases, the CAPADR region will likely experience more synchronous demand and supply shocks over time. If completed, the adhesion of the rest of the countries to the Customs Union among Guatemala, Honduras, and El Salvador will translate into increased intra-regional trade that may foster such a relationship \cite{Duran2019}. However, specialized industries may lead to desynchronization. In this sense, policy and decision-makers must be aware of any detrimental specialization patterns stemming from promoted regional integration strategies, such as regional value chains (e.g., some countries as producers of intermediate inputs and others as producers of high value-added, final goods), which could also induce wider dissimilarities in terms of economic structure. The implementation of a common market with free labor and capital mobility is also a pending topic.\\

Macroeconomic policy coordination is crucial for further integration and currency unification. Similar fiscal and monetary policies tend to synchronize comovements in demand disturbances as they lead to correlated fiscal and monetary shocks \cite{Samba2019}. Even though CAPADR countries have undertaken some efforts in terms of policy cooperation, differences are still the rule rather than the exception (for example, exchange rate systems). In the long run, the region must establish explicit compulsory targets on inflation, nominal exchange rate, interest rate, fiscal deficit, and public debt (Maastricht-type requirements) to ensure minimal conditions for macroeconomic convergence for an economic union to become feasible and sustainable \cite{Alvarado2019}.\\

Finally, it is worth highlighting the importance of institutional and political commitment for the success of economic and monetary unions within CAPADR economies. Involved governmental instances must work alongside regional institutions toward the consolidation of each integration stage and policy instrumentation. States must also commit to collective action, shared objectives, and political consensus. By doing so, CAPADR countries will be better prepared to tackle common challenges (e.g., extreme poverty, inequality, and climate change risks, among others) and gain public support for the integration efforts.

\section{Concluding remarks}

Since 1960, Panama, Dominican Republic, and Central American countries have carried out an ongoing process of economic integration. In particular, the coordination efforts have become more visible in the light of the negotiation and launching of trade agreements, the consolidation of the intern market, and the strengthening of policy coordination. In this context, it is natural to argue whether CAPADR may move beyond economic unification to adopt a single currency system.  \\

In this paper, we investigate if CAPADR countries are suitable to form an OCA. We operationalize such feasibility as the degree of symmetry between country-specific supply and demand shocks and the level of business cycle synchronization. For the extraction of structural shocks, we estimate seven SVAR models by using the long-run identification method proposed by \citeA{Bayoumi1992}, based on the \citeA{Blanchard1989} decomposition method. We then go on to construct two indicators of dispersion based on the aggregate disturbances series and compute the cost of inclusion in terms of country-specific contributions to synchronization, as pointed out by \citeA{CrespoCuaresma2013}.\\

The examination of the computed correlation coefficients suggests that asymmetric supply and demand shocks tend to predominate among CAPADR economies. Nevertheless, we identify one group of countries (Dominican Republic, Honduras, Panama, El Salvador) that may benefit from closer policy coordination. However, the costs of undertaking currency unification are likely to be high. As for the indicators of dispersion, we find that supply and demand shocks have become more synchronous over time, accounting for trend decreases of 28.5\% and 32.4\% between January 2020 and October 2010, respectively. Furthermore, we also reckon that, on average, most countries are origins of cyclical divergence, so that forming an OCA would result in less synchronous business cycles and a costly loss of monetary sovereignty. Altogether, we conclude that the establishment of an OCA within CAPADR economies lacks feasibility.\\

The identification of determinants of business cycle synchronization is beyond the scope of this paper. Future research can shed light on the particular variables that may strengthen the convergence process shown by our indicators of dispersion. Besides economic factors, it is worth emphasizing the importance of political and institutional commitment to support closer economic integration across the CAPADR region, particularly if these countries become more symmetric in terms of macroeconomic shocks and more synchronized in terms of business cycles so that the idea of an OCA turns out to be appropriate. Meanwhile, the region should focus on pushing forward economic union.

\nocite{CMCA2021a} \nocite{CMCA2021b} \nocite{WorldBank2021}

\newpage
\cleardoublepage
\phantomsection

\bibliographystyle{apacite}
\begingroup
\urlstyle{same}
\setlength{\bibitemsep}{10pt}
\setstretch{1}
\bibliography{oca_ca}
\endgroup

\newpage
\cleardoublepage
\phantomsection
\addcontentsline{toc}{section}{Appendix}
\appendix
\setcounter{table}{0}
\renewcommand{\thetable}{A\arabic{table}}

\section*{Appendix}
\subsection*{Cointegration results}
\begin{table}[h!]
	\centering
	\caption{Results of Johansen cointegration test}
	\label{ta1}
	\footnotesize
	\begin{threeparttable}
	\begin{tabular}{@{}ll*{2}{S[table-format = 2.3]} l*{2}{S[table-format = 2.3]}@{}}
		\toprule \toprule
		\multicolumn{1}{c}{\multirow{2}{*}{Country}} & \multicolumn{1}{c}{\multirow{2}{*}{Hypothesis}} & \multicolumn{2}{c}{Trace} &  & \multicolumn{2}{c}{Maximum eigenvalue} \\ \cmidrule(lr){3-4} \cmidrule(l){6-7} 
		\multicolumn{1}{c}{} & \multicolumn{1}{c}{} & \multicolumn{1}{c}{Statistic} & \multicolumn{1}{c}{Critical value} &  & \multicolumn{1}{c}{Statistic} & \multicolumn{1}{c}{Critical value} \\ \midrule
		\multirow{2}{*}{CRI} & $H(0): r = 0$ & 19.956 & 25.32 &  & 15.816 & 18.96 \\
		& $H(1): r \leq 1$ & 4.139 & 12.25 &  & 4.139 & 12.25 \\ \cmidrule(r){1-4} \cmidrule(l){6-7} 
		\multirow{2}{*}{DOM} & $H(0): r = 0$ & 24.434 & 25.32 &  & 17.027 & 18.96 \\
		& $H(1): r \leq 1$ & 7.406 & 12.25 &  & 7.406 & 12.25 \\ \cmidrule(r){1-4} \cmidrule(l){6-7} 
		\multirow{2}{*}{GTM} & $H(0): r = 0$ & 14.570 & 25.32 &  & 8.298 & 18.96 \\
		& $H(1): r \leq 1$ & 6.272 & 12.25 &  & 6.272 & 12.25 \\ \cmidrule(r){1-4} \cmidrule(l){6-7} 
		\multirow{2}{*}{HND} & $H(0): r = 0$ & 18.605 & 25.32 &  & 12.111 & 18.96 \\
		& $H(1): r \leq 1$ & 6.494 & 12.25 &  & 6.494 & 12.25 \\ \cmidrule(r){1-4} \cmidrule(l){6-7} 
		\multirow{2}{*}{NIC} & $H(0): r = 0$ & 17.261 & 25.32 &  & 10.030 & 18.96 \\
		& $H(1): r \leq 1$ & 7.231 & 12.25 &  & 7.231 & 12.25 \\ \cmidrule(r){1-4} \cmidrule(l){6-7} 
		\multirow{2}{*}{PAN} & $H(0): r = 0$ & 23.446 & 25.32 &  & 13.978 & 18.96 \\
		& $H(1): r \leq 1$ & 9.469 & 12.25 &  & 9.469 & 12.25 \\ \cmidrule(r){1-4} \cmidrule(l){6-7} 
		\multirow{2}{*}{SLV} & $H(0): r = 0$ & 21.186 & 25.32 &  & 13.100 & 18.96 \\
		& $H(1): r \leq 1$ & 8.086 & 12.25 &  & 8.086 & 12.25 \\ \bottomrule
\end{tabular} 
\begin{tablenotes}
	\footnotesize
	\item[a]\emph{Note.} Critical values at the 5\% significance level. Specification with a linear trend.
	\item[ ]\emph{Source.} Own calculation.
\end{tablenotes}
\end{threeparttable}
\end{table}
\subsection*{Weights for indicators of dispersion}
\begin{table}[h!]
	\centering
	\caption{Country weights for the computation of dispersion indicators.}
	\label{ta2}
	\footnotesize
	\begin{threeparttable}
	\begin{tabular}{@{}llllllll}
		\toprule \toprule
		Year & \multicolumn{1}{c}{CRI} & \multicolumn{1}{c}{DOM} & \multicolumn{1}{c}{GTM} & \multicolumn{1}{c}{HND} & \multicolumn{1}{c}{NIC} & \multicolumn{1}{c}{PAN} & \multicolumn{1}{c}{SLV} \\ \midrule
		2009 & 0.155 & 0.242 & 0.216 & 0.083 & 0.055 & 0.156 & 0.094 \\
		2010 & 0.155 & 0.250 & 0.211 & 0.082 & 0.054 & 0.157 & 0.091 \\
		2011 & 0.154 & 0.245 & 0.209 & 0.081 & 0.055 & 0.166 & 0.090 \\
		2012 & 0.155 & 0.241 & 0.206 & 0.080 & 0.056 & 0.174 & 0.089 \\
		2013 & 0.152 & 0.242 & 0.205 & 0.079 & 0.056 & 0.179 & 0.087 \\
		2014 & 0.150 & 0.248 & 0.204 & 0.078 & 0.056 & 0.179 & 0.085 \\
		2015 & 0.149 & 0.252 & 0.203 & 0.077 & 0.056 & 0.181 & 0.082 \\
		2016 & 0.148 & 0.258 & 0.199 & 0.077 & 0.056 & 0.181 & 0.081 \\
		2017 & 0.148 & 0.259 & 0.197 & 0.077 & 0.056 & 0.184 & 0.079 \\
		2018 & 0.146 & 0.267 & 0.196 & 0.077 & 0.052 & 0.184 & 0.078 \\
		2019 & 0.144 & 0.272 & 0.197 & 0.077 & 0.049 & 0.183 & 0.078 \\
		2020 & 0.149 & 0.274 & 0.210 & 0.076 & 0.052 & 0.163 & 0.078 \\ \bottomrule
	\end{tabular}
\begin{tablenotes}
	\footnotesize
	\item[ ]\emph{Source.} Own calculation based on World Bank data.
\end{tablenotes}
\end{threeparttable}
\end{table}

\end{document}